\documentclass[%
 reprint,
 a4paper,
superscriptaddress,
showpacs,preprintnumbers,
nobibnotes,
 amsmath,amssymb,
 aps,
prstab,
floatfix,
]{revtex4-1}

\usepackage{graphicx}
\usepackage{dcolumn}
\usepackage{bm}
\usepackage{hyperref}
\hypersetup{colorlinks=true,citecolor=blue}

\begin{document}
\preprint{APS/XX-XXX}
\title{Low emittance muon accelerator studies with production from positrons
   on target}

\author{M.~Boscolo}\email{manuela.boscolo@lnf.infn.it}
\affiliation{INFN-LNF, Via E. Fermi 40, 00044 Frascati, Rome, Italy}
\author{M.~Antonelli}
\affiliation{INFN-LNF, Via E. Fermi 40, 00044 Frascati, Rome, Italy}
\author{O.R.~Blanco-Garc\'ia}
\affiliation{INFN-LNF, Via E. Fermi 40, 00044 Frascati, Rome, Italy}
\author{S.~Guiducci}
\affiliation{INFN-LNF, Via E. Fermi 40, 00044 Frascati, Rome, Italy}
\author{S.~Liuzzo}
\affiliation{ESRF, 71 avenue des Martyrs, 38000 Grenoble, France} 
\author{P.~Raimondi}
\affiliation{ESRF, 71 avenue des Martyrs, 38000 Grenoble, France} 
\author{F.~Collamati}
\affiliation{INFN-Rome, Piazzale A. Moro 2, 00185 Rome, Italy}

\date{\today}

\begin{abstract}
A new scheme to produce very low emittance muon beams using a positron beam of about 45~GeV interacting on electrons on target is presented.
 One of the innovative topics to be investigated is the behaviour of the positron beam stored in a low emittance  ring  with a thin target, that is directly inserted in the ring chamber to produce muons.  Muons can be immediately collected at the exit of the target and transported to two $\mu^+$ and $\mu^-$ accumulator rings and then accelerated and injected in muon collider rings. We focus in this paper on the simulation of the e$^+$ beam interacting with the target, the effect of the target on the 6-D phase space and the optimization of the e$^+$  ring design  to maximize the energy acceptance. We will investigate the performance of this scheme, ring  plus target system, comparing different multi-turn simulations.
The source is considered for use in a multi-TeV  collider in ref.~\cite{DAgnolo:2268140}. 



\end{abstract}

\maketitle


\section{\label{sec:level1}Introduction}
Muon beams are customarily obtained via $K/\pi$ decays produced in proton interaction on target.
A complete design study using this scheme, including the muon cooling system has been performed by the Muon Accelerator Program~\cite{map,map_ipac14}.
  In this paper we will investigate the possibility to produce low emittance muon beams from a novel approach, using 
electron-positron collisions at a centre-of-mass energy just above the $\mu^{+}\mu^{-}$  production
threshold with minimal muon energy spread, corresponding to the direct annihilation of
approximately 45~GeV positrons and atomic electrons in a thin target, O(0.01$\sim$ radiation lengths).
 Concept studies on this subject are reported in Refs.~\cite{NIM,IPAC17}.
 A feasibility study of a muon collider based on muon electro-production has been studied in Ref.~\cite{Barletta:1993rq}.
The most important key properties of the muons produced by the positrons on target are: 
\begin{itemize}
\setlength\itemsep{0em}
  \item the low and tunable muon momentum in the centre of mass frame;
  \item large boost, being about $\gamma\sim$200.
\end{itemize}
These characteristics result in the following advantages:
  \begin{itemize}
\setlength\itemsep{0em}
  \item the final state muons are highly collimated and have very small emittance, overcoming the
  need to cool them;
  \item the muons have an average laboratory lifetime of about 500~$\mu$s at production.
  \item the muons are produced with an average energy of 22~GeV easing the acceleration scheme.
\end{itemize}

The use of a low emittance positron beam on target  allows the production of low emittance muon beam and, 
 unlike previous designs, muon cooling would not be necessary. \\
The very small emittance could enable high luminosity with smaller muon fluxes, 
reducing both the machine background in the experiments  and more importantly the activation risks due to neutrino interactions. 
   \begin{figure}[!htb]
   \centering
 \includegraphics*[width=230pt]{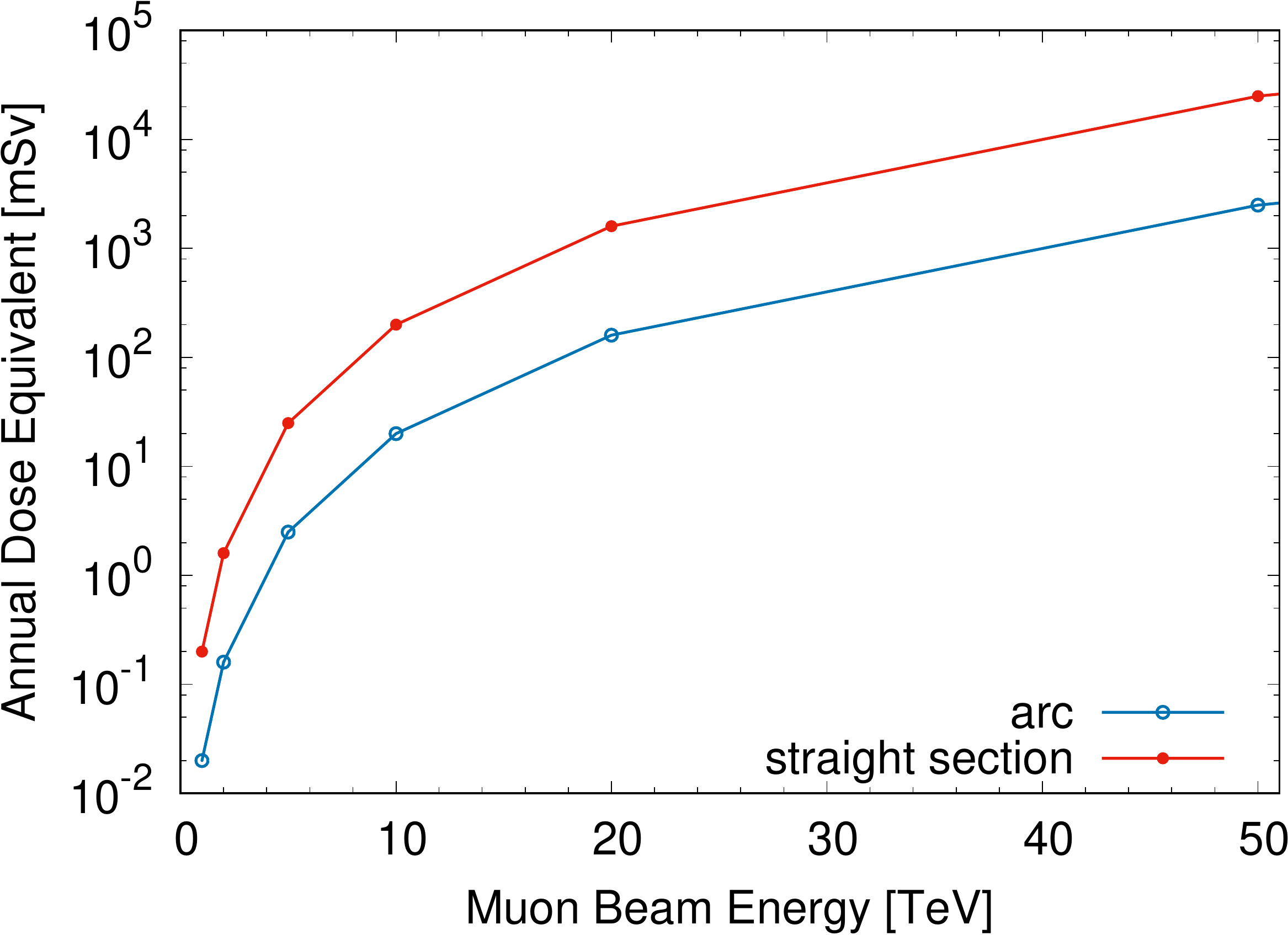}
   \caption{Dose equivalent due to neutrino radiation  at a collider depth of 100~m  for a muon rate of $3\cdot 10^{13}$~s$^{-1}$ as taken from~\cite{rolandi_silari}. Contributions from 
  the straight section and from the arc   are shown in red and blue, respectively.}
   \label{radiolimit}
\end{figure}

Figure~\ref{radiolimit} shows the dose equivalent due to neutrino radiation as taken from~\cite{rolandi_silari}. It shows 
an increase of the dose equivalent with the muon beam energy posing a severe limit on the centre of mass energy reach of the muon collider.
It has been obtained for a muon rate of $3 \cdot 10^{13}$s$^{-1}$ and a collider depth of 100~m.
Such a muon rate is problematic for muon collider operations  above about 
6~TeV centre of mass energy~\cite{map}. 
Ref.~\cite{map}
chose a collider depth of 500 m for a 6 TeV collider scenario. 

Figure~\ref{radiolimit} and Ref.~\cite{rolandi_silari} provides a first crude approximation of the dose effect, and a more detailed calculation with consideration of mitigation strategies is needed for precise limit.  To enable higher centre of mass energies the muon rate has to be reduced, thus competitive luminosity performance must be obtained by reducing the beam emittance. 

The current muon collider design studied by MAP~\cite{map} foresees a normalized emittance of $25~\rm \mu m$  with
 a luminosity of about 10$^{35}$~cm$^{-2}$ s$^{-1}$ at 6~TeV. The muons per bunch are $2 \cdot 10^{12}$ and the muon production rate is $3 \cdot 10^{13}$s$^{-1}$.
The aim of the study is to obtain comparable luminosity performances with lower fluxes and smaller emittances allowing operation at a higher centre of mass energy.
One possibility is to accommodate the accelerator complex in the CERN area in existing tunnels. A preliminary proposal for a 14~TeV CERN muon collider
has been described in ref~\cite{DAgnolo:2268140}.

\subsection{Muon Production}
The cross section for continuum muon pair production $e^{+}e^{-}\rightarrow\mu^{+}\mu^{-}$ has a maximum value of about 1~$\mu$b at $\sqrt{s} \sim$~0.230~GeV.
In our proposal these values of  $\sqrt{s}$  can be obtained from fixed target interactions with a positron beam energy of 
 \begin{equation}
    E(e^{+})~\approx  \frac{s}{2~m_{e}} \approx 45 ~\hbox{GeV}    
 \end{equation}
where $m_{e}$  is the electron mass, with a boost of 
 \begin{equation}
\gamma \approx  \frac{E(e^{+})}{\sqrt{s}} \approx \frac{\sqrt{s}}{2~m_{e}}  \approx 220.
 \end{equation}
The maximum scattering angle of the outgoing muons $\theta_{\mu}^{max}$   depends on $\sqrt{s}$. 
In the approximation of $\beta_{\mu}=1$,
\begin{equation}
  \theta_{\mu}^{max}= \frac{4~m_{e}}{s} \sqrt{\frac{s}{4}-m_{\mu}^{2} }
    \end{equation}
     where $\beta_{\mu}$ is the muon velocity.
 Muons produced with very small momentum
in the rest frame are well contained in a cone of about  $5\cdot 10^{-4}$ rad for   $\sqrt{s}$=0.212 GeV,  the cone
size increases to $\sim 1.2\cdot 10^{-3} $  rad at  $\sqrt{s}$=0.220 GeV.
The difference between the maximum and the minimum energy of the muons produced
at the positron target ($\Delta E_{\mu}$) also depends on  $\sqrt{s}$,  and with the  $\beta_{\mu}=1$ approximation we get:
\begin{equation}
\Delta E_{\mu} =\frac{\sqrt{s}}{2m_{e}} \sqrt{\frac{s}{4}-m_{\mu}^{2}}
\end{equation}

The $RMS$ energy distribution of the muons increases with $\sqrt{s}$, from about
1~GeV at $\sqrt{s}$=0.212 GeV to ~3 GeV at  $\sqrt{s}$=0.220~GeV. 

\subsection{Target options}
The number of $\mu^{+}\mu^{-}$ pairs produced per positron bunch on target is:
\begin{equation}
\label{eq:nmu}
  n(\mu^{+} \mu^{-}) = n^{+}~\rho^{-}~L~\sigma(\mu^{+} \mu^{-})      
\end{equation}
where $n^{+}$ is the number of positrons in
the bunch, $\rho^-$ is the electron density in the medium, $L$ is
the thickness of the target, and $\sigma(\mu^{+} \mu^{-})$ is the muon pairs production cross section.
The dominant process in $e^+$ $e^-$ interactions  at these energies
is  collinear radiative Bhabha scattering, with a cross section of about $\sigma_{rb} \approx \rm 150~mb$ for a positron energy loss larger than 1\%.
This sets the value of the positron beam interaction length for a given pure electron target density value. 
So, it is convenient to use targets with thickness of at most one interaction length,
corresponding to 
$L = 1/(\sigma_{rb}~\rho^-)$:
\begin{equation}
(\rho^-~L)_{max} =  1/\sigma_{rb}\approx 10^{25}  \hbox{cm}^{-2}
\end{equation}
The ratio of the muon pair production cross section to the radiative
bhabha cross section determines the maximum value of the {\it muon conversion efficiency} $\rm{eff}(\mu^+\mu^-)$
that can be obtained with a pure electrons target.
In the following we will refer to $\rm{eff}(\mu^+\mu^-)$ defined as the ratio of the number of produced $\mu^+\mu^-$ pairs
to the number of the incoming positrons.
One can easily see that  the upper limit of $\rm{eff}(\mu^+\mu^-)$  is of the order of $10^{-5}$, so that:
\begin{equation}
 n(\mu^+ \mu^-)_{max}\approx n^+~10^{-5}.
\end{equation}

Electromagnetic interactions with nuclei are dominant in conventional targets. 
In addition, limits are present for the target thickness 
not to increase the muon beam emittance $\epsilon_{\mu}$.
Assuming a uniform distribution in the transverse $x-x'$ plane the emittance contribution due to the target thickness is 
proportional to the target length and to the maximum scattering angle of the outgoing muons:
\begin{equation}
\epsilon_{\mu}\propto{L~(\theta_{\mu}^{max})^{2}}
\end{equation}
The number of $\mu^+\mu^-$ pairs produced per crossing has the form given by equation~\ref{eq:nmu},  with
\[
\rho^-=N_A/A~\rho~Z
\]
where $Z$ is the atomic number, $A$ the mass number, $N_A$  Avogadro's constant, and $\rho$ the material density.
In addition, the multiple scattering contributes to the emittance increase according to:
\begin{equation}
\theta_{\rm{r.m.s}} \sim \frac{13.6~\rm{MeV} }{\beta cp}\sqrt{x/X_0}[1+0.038~\ln(x/X_0)]
\label{MSscat}
\end{equation}
where $p$ and $\beta c$ are the positron velocity and momentum, respectively, $x/X_0$ is the target thickness in radiation lengths.
 Similarly: 
\[
x_{\rm{r.m.s.}}\sim \theta_{\rm{r.m.s}} ~0.5~L~\sqrt{3}.
\]
The bremsstrahlung process governs the positron beam degradation in this case and it scales with the radiation length. Roughly, 
the cross section per atom increases by a factor (Z+1) with respect to the case of a pure electron target. 

On one side to minimize the emittance there is the need of a small length $L$ (thin target),
on the other side compact materials have typically small radiation lengths causing
an increase of the emittance due to multiple scattering and a fast positron
beam degradation due to bremsstrahlung. 
Positron interactions on different targets have
been studied with {\texttt{GEANT4}}~\cite{geant}. The results show that the optimal target has to be thin and not too heavy~\cite{NIM}. 
 Carbon (C), Beryllium (Be) and Liquid Lithium (LLi) have the best performances.
 
\begin{figure}[!htb]
   \centering
   \includegraphics*[width=230pt]{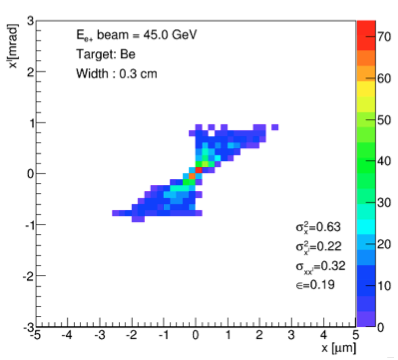}
   \caption{Transverse phase space distribution of muons at the  target exit (GEANT4). The positron beam impinging on the 3~mm Be target has  negligible emittance.}
   \label{f2}
\end{figure}

 A 3~mm Beryllium target provides a muon conversion efficiency $\rm eff(\mu^+\mu^-)$ of about $0.7\cdot 10^{-7}$.
 The muon transverse phase space at the target exit, as produced by a 45~GeV positron beam, is shown in Figure~\ref{f2}, assuming negligible emittance of the positron beam. 
A transverse emittance as small as 
$0.19\cdot 10^{-9}$~m-rad is observed for the outgoing $\sim$22~GeV muons; it represents the ultimate value of emittance that can be obtained with such a scheme.\\
 The very low muon production efficiency, due to the low value of the production cross section, makes convenient  a scheme 
 where  positrons are recirculated after the interaction on a thin target.
 A 3~mm Beryllium target is considered in our studies.\par
\begin{figure}[!htb]
   \centering
   \includegraphics*[width=230pt]{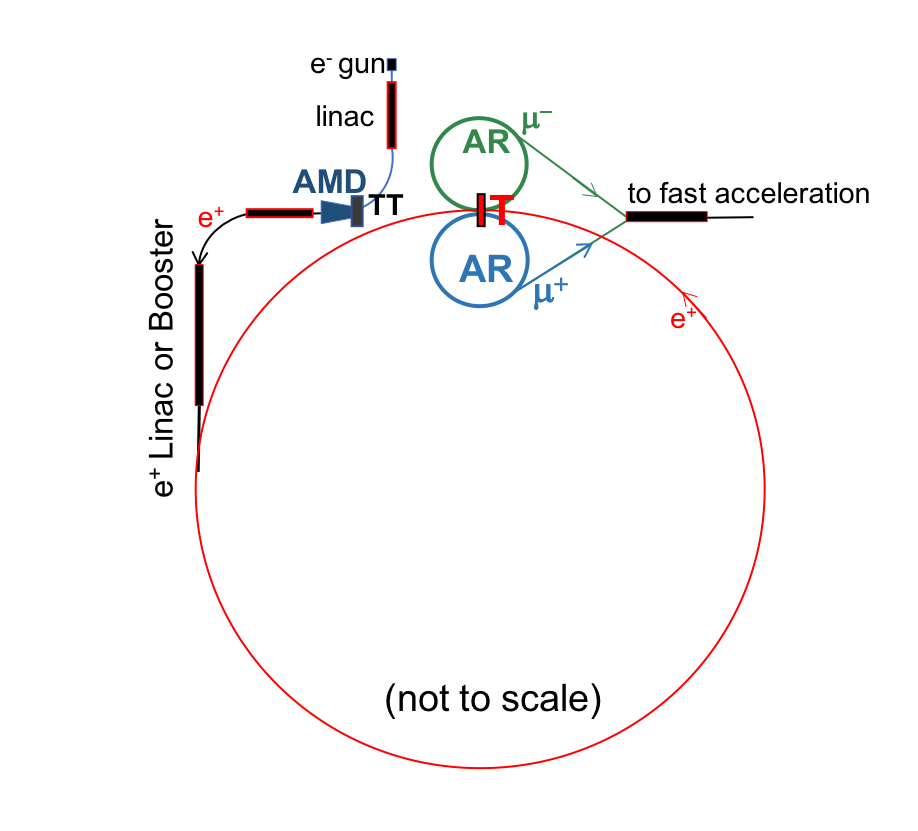}
   \caption{Schematic layout for low emittance  muon beam production: positron source with adiabatic matching device (AMD), 
   $e^{+}$  ring plus target (T)  for $\mu^{+}$-$\mu^{-}$ production, two $\mu^{+}$-$\mu^{-}$ accumulator rings (AR), and fast acceleration section  to be followed by the muon collider.}
   \label{scheme}
\end{figure}

 A preliminary layout of  Low EMittance Muon Accelerator (LEMMA)  is shown in Figure~\ref{scheme}.  
This scheme foresees   muons  produced  by the interaction of positrons  circulating in a storage ring on target  at $\sim$ 22~GeV, 
  then  muons are accumulated in isochronous rings with a circumference  of $\sim$60~m with 13~T dipoles.
 In addition to muons, high intensity and  high energy photons are produced  from the interaction of the positron beam with the target.
The possibility of using these photons for an embedded  positron source is extremely appealing and will be briefly described in Section~\ref{posiso}.
However, a solution to transform the temporal structure of produced positrons  to be used for the ring injection 
has not been found yet.

   This innovative scheme has many key topics to be investigated:
   \begin{itemize}
\item   a low emittance high energy acceptance 45~GeV positron ring,
\item  O(100~kW) class target,
\item  high momentum acceptance muon accumulator rings, 
\item high rate positron source.
\end{itemize}

A first design of a 45~GeV positron ring with low emittance and high momentum acceptance will be described in the following.
The effects on beam parameters due to the target insertion will be analysed from the point of view of the positron  beam lifetime and its degradation.  
In particular, an attempt to the control the positron emittance growth will be described. Our goal is to preserve as much as possible the ultimate value of normalized emittance, 
$\epsilon_\mu$=40~nm, obtained for a 3~mm Be target and to obtain a positron beam lifetime of about 250 turns. Additional emittance dilution effects due to the muon accumulation
 are the subject of future studies.  
  
\section{Positron storage ring}
\label{optics}
A 45~GeV low emittance positron storage ring has been designed and the effect of the target on the beam properties has been studied.
The main processes affecting the beam sizes in the target are bremsstrahlung and multiple Coulomb scattering. 
The effects of these two contributions have been studied separately, finding that the best location for the target corresponds to a low-$\beta$, dispersion-free region.\par
\begin{figure}[!htb]
   \centering
   \includegraphics*[width=230pt,angle=0]{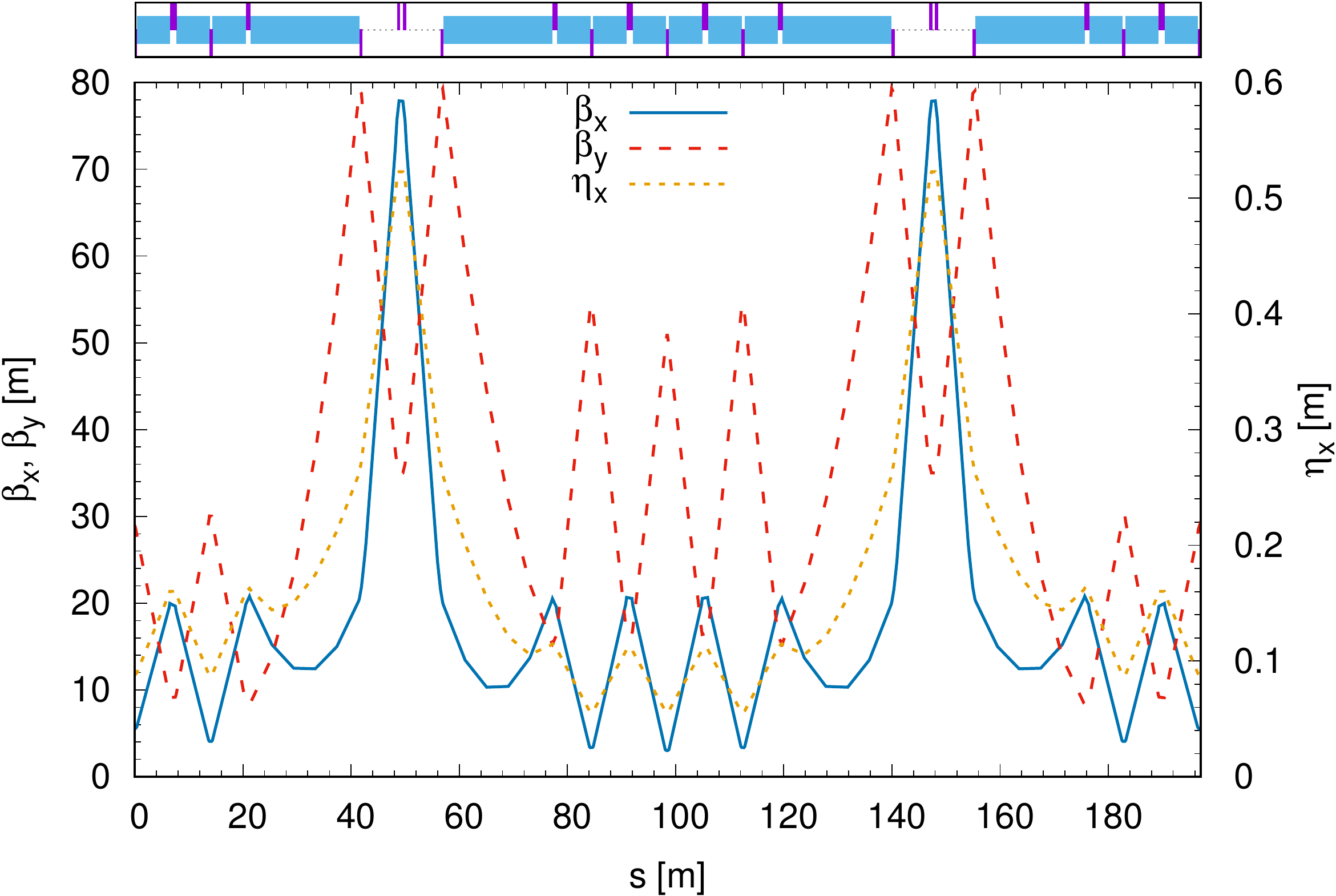}
   \caption{Optical functions of one cell of the 45~GeV e$^+$ ring. $\beta_{x}$ and $\beta_{y}$ are shown in blue and dotted red, respectively; horizontal dispersion function $\eta_{x}$ is plotted in yellow with
   values  on the right y axis.}
   \label{cell}
\end{figure}
\begin{figure}[!htb]
   \centering
 \includegraphics[width=9.5cm,angle=-90]{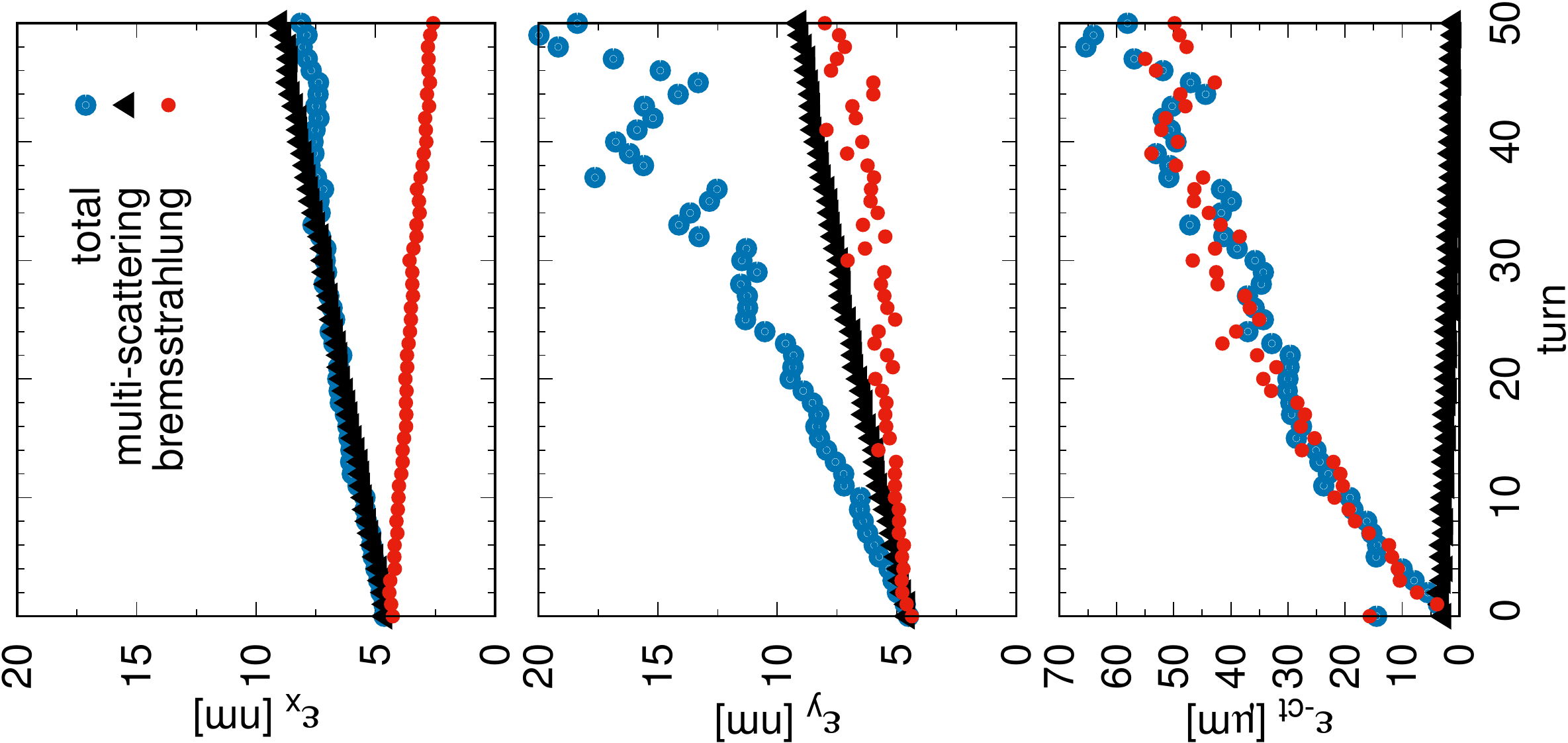}
   \caption{Horizontal (upper plot), vertical (middle plot) and longitudinal (lower plot) emittance degradation as a function of turns is shown (blue dots) for a 3~mm thick Beryllium target with {$\beta^*_x$=$\beta^*_y$}=0.5~m. Separated contributions from multiple scattering (black triangles) and  bremsstrahlung (red dots)  are also shown. The horizontal emittance is dominated by multiple-scattering; the longitudinal emittance by bremsstrahlung.}
   \label{horbeam}
\end{figure}

The ring is composed of 32~lattice cells of 197~m each for a total length of 6~km with the parameters shown in Table~\ref{tab:ringpara}. The cell shown in Figure~\ref{cell} is based on the Hybrid Multi-Bend Achromat (HMBA)~\cite{doi:10.1080/08940886.2014.970931} to minimize emittance, while keeping large momentum and dynamic acceptance. The maximum dipole field is 0.26~T, the filling factor is 77\%. The maximum quadrupole, sextupole and octupole gradients are 110~T/m, 340~T/m$^2$ and 5900~T/m$^3$ respectively. The horizontal and vertical phase advance, in $2\pi$ units, between sextupoles at the peak of the horizontal dispersion is 1.5 and 0.5 respectively. Sixty-four RF cavities have been considered, each RF cavity is 5.4~m long and composed by 9 RF cells of about 7~MV/m accelerating gradient. Transverse unnormalized emittance is 5.73 $\times ~10^{-9}$ and longitudinal emittance 3~$\mu m$; then, as the positron beam passes through the target, emittance  is increased as discussed later, see Figure~\ref{horbeam}. Longitudinal emittance is increased due to a large elongation effect due to the bremsstrahlung effect in the passage of the target.\par

\begin{table}[t]
  \centering
  \caption{Positron ring parameters.
     \label{tab:ringpara}}
  \vspace{0.05cm}
  \begin{tabular}{*{3}{l}}
    \hline\hline
    \noalign{\vspace{0.05cm}}
    Parameter & Units &  \\
    Energy & GeV & 45 \\
    Circumference (32 ARCs, no IR) & m & 6300.960 \\
    Geometrical emittance x, y & m & 5.73 $\times ~10^{-9}$ \\
    Bunch length   & mm & 3  \\
    Beam current & mA & 240 \\	
    RF frequency & MHz & 500  \\
    RF voltage & GV & 1.15 \\	
    Harmonic number & \#  & 10508 \\
    Number of bunches & \# & 100 \\    
    N. particles/bunch  &\#   & 3.15 $\times  10^{11}$ \\    
    Synchrotron tune  & & 0.068 \\       
    Transverse damping time & turns & 175 \\
    Longitudinal damping time & turns & 87.5 \\
    Energy loss/turn & GeV & 0.511 \\         
    Momentum compaction &  & 1.1 $\times  10^{-4}$ \\
    RF acceptance & \% & $\pm$  7.2\\      
    Energy spread & dE/E & 1 $\times  10^{-3}$ \\
    SR power & MW & 120 \\   
    \hline\hline
  \end{tabular}
\end{table}
The effect of the target on the positron beam has been studied for different locations in the cell for different materials and thicknesses. This simulation has been divided in two parts: particle tracking and positron interaction with target. Particle tracking in the ring is performed with Accelerator Toolbox~(AT)~\cite{AT} and MAD-X PTC~\cite{MADX}, while positron interaction with the target is studied  with either  GEANT4 or FLUKA~\cite{FLUKA}.\par
The cycle starts by the generation of a particle distribution from the equilibrium emittances of the ring. These particles are tracked through the ring for one turn and the particle distribution is modified according to simulations to account for the passage through the target. 
Radiation damping is included in the simulation.
Single turn tracking and target interaction are then repeated for a given number of turns.\par

Coulomb scattering in the target changes the beam divergence and the beam size. 
The multiple Coulomb scattering contribution is normally distributed and uncorrelated with the beam, therefore it only depends on the target. It can be estimated as in eq.~\ref{MSscat}.
As the beam passes through several times, the contribution from multiple scattering~(MS) increases as 
\begin{equation}
\sigma_{x',y'}(MS) = \sqrt{N} ~\theta_{\rm{r.m.s}} 
\label{eq:10}
\end{equation}
where $N$ is the turn number. Therefore, the beam divergence per turn is given by 
\begin{equation}
\sigma_{x',y'}(N) = \sqrt{\sigma^{2}_{x',y'}(MS)+\sigma^{2}_{x',y'}(0)}
\label{eq:11}
\end{equation}
where $\sigma_{x',y'}(0)$ is the unperturbed beam divergence in the two transverse planes.\par
Similar effects have been studies for ionization cooling~\cite{ionization_cooling} and for ion stripping~\cite{strip}. 

While the contribution to the divergence from multiple scattering is completely determined by the target, the contribution to the beam size is expected to be proportional to the $\beta$-function at the target location.\par
\begin{figure}[!htb]
   \centering
 \includegraphics[height=8.cm,angle=-90]{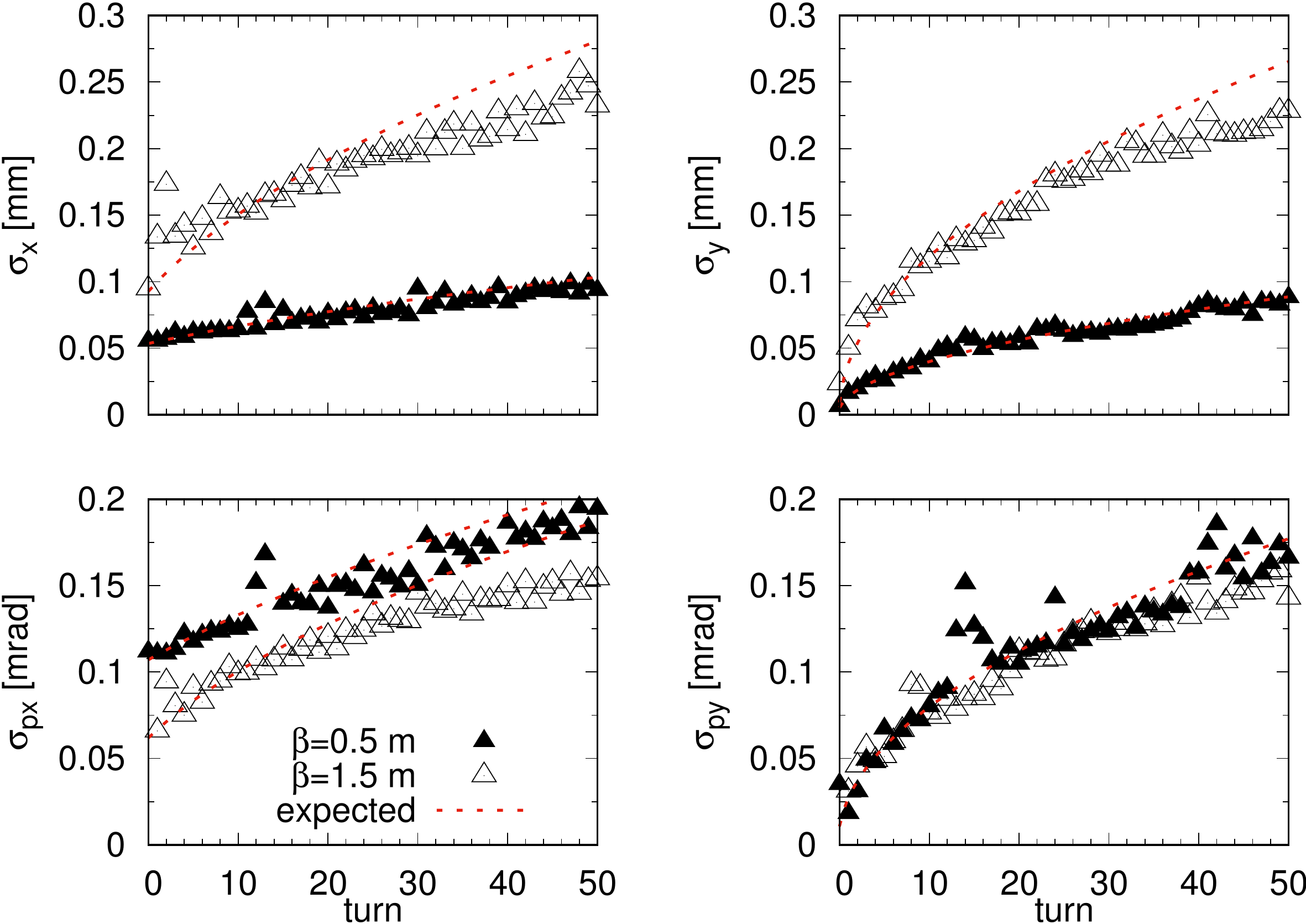}
 \caption{Positron beam transverse  evolution  vs machine turns due to multiple scattering with a 3~mm Be target.  Triangles are PTC tracking results, for two $\beta^{*}$ values at the IP (0.5 and 1.5~m), while dashed lines are the expected behaviour  from equations ~\ref{eq:10} and~\ref{eq:11}. An emittance ratio of 1\% has been considered for this simulation.}
   \label{multiscat}
\end{figure}

At a waist the beam width after $N$ machine turns is given by
\begin{equation}
\sigma_{x,y}(N) = \beta_{x,y}~\sigma_{x',y'}(N).
\end{equation} 
This explains how a strong reduction of the beam size growth is obtained by placing the target in a low-$\beta$ location, such as the interaction point in a collider ring. To visualize this effect, Figure~\ref{multiscat} shows the beam size and divergence for a positron beam interacting with a 3~mm thick Beryllium target at each turn. In the case of $\beta$ value at the target (waist) equal to 1.5~m the simulated increase in the beam divergence is $\theta_{\rm{r.m.s}}\sim$25~$\mu$rad at each single pass. Lower $\beta$ functions evidently reduce this effect.

\par
 Minimization of the emittance growth is obtained  by placing the target in a beam waist and setting the angular contribution from the multiple scattering similar to or smaller  than the beam divergence~\cite{HWANG2014153}. 
This matching reduces  beam filamentation~\cite{Mohl:1005037}. \par
\begin{figure}[!htb]
   \centering
   \includegraphics*[width=8cm]{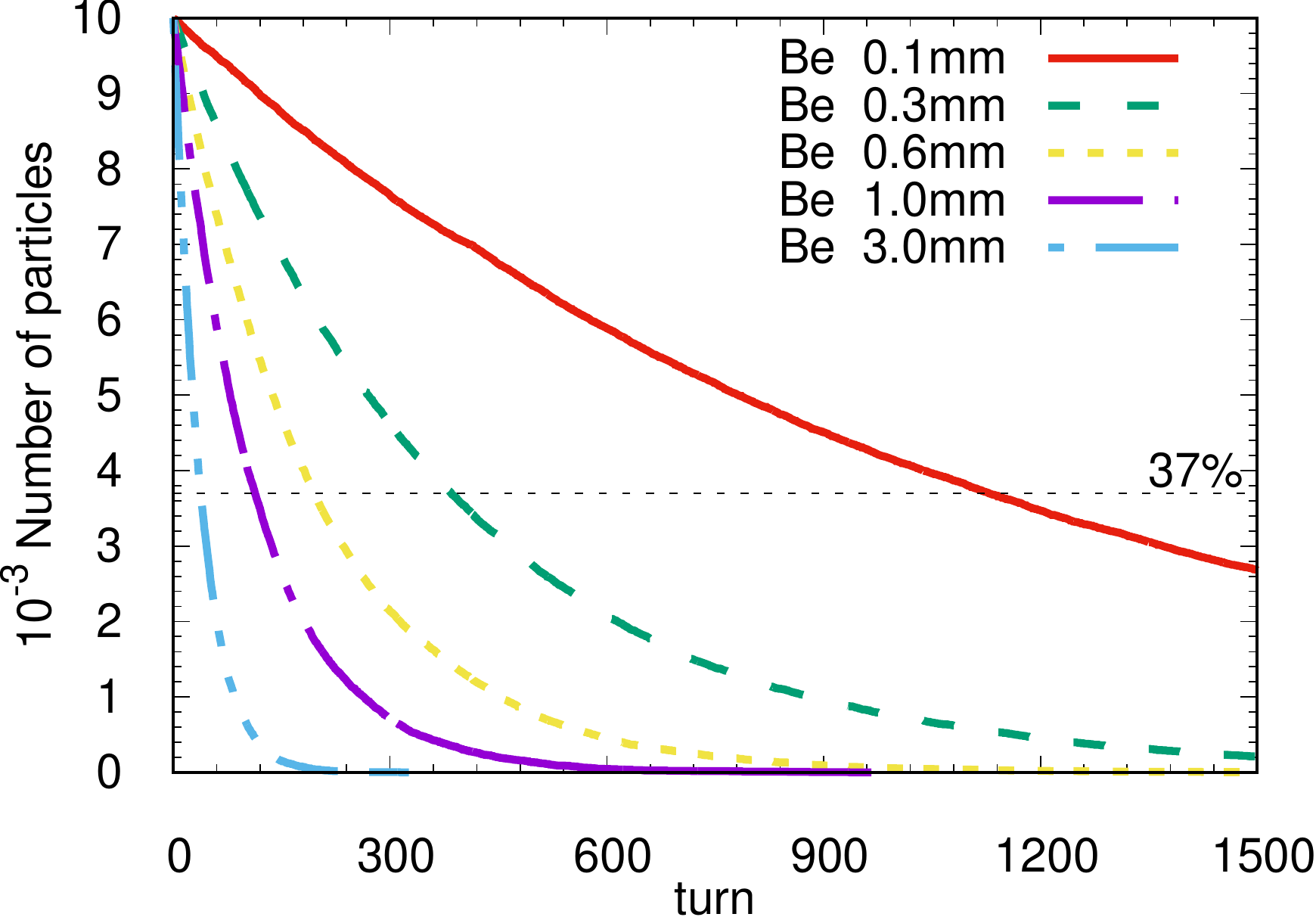}
   \caption{Number of e$^+$  versus machine turns for various Be target thicknesses (MAD-X PTC). For a 3~mm Be target (light blue line) lifetime is about 40 turns.}
   \label{lifetime}
\end{figure}

Bremsstrahlung in the target causes the particles to lose energy, degrading the beam emittance and reducing its lifetime. For a  3~mm thick target of Beryllium we obtain, from tracking with MAD-X PTC, a life time between 37 and 40 turns, equivalent to 0.8~ms, in good agreement with AT results. Figure~\ref{lifetime} shows the number of particles per turn for several target thicknesses from which it is concluded that the beam lifetime and target thickness are inversely proportional, as expected.\par

\begin{figure}[!htb]
   \centering
   \includegraphics*[width=8.cm]{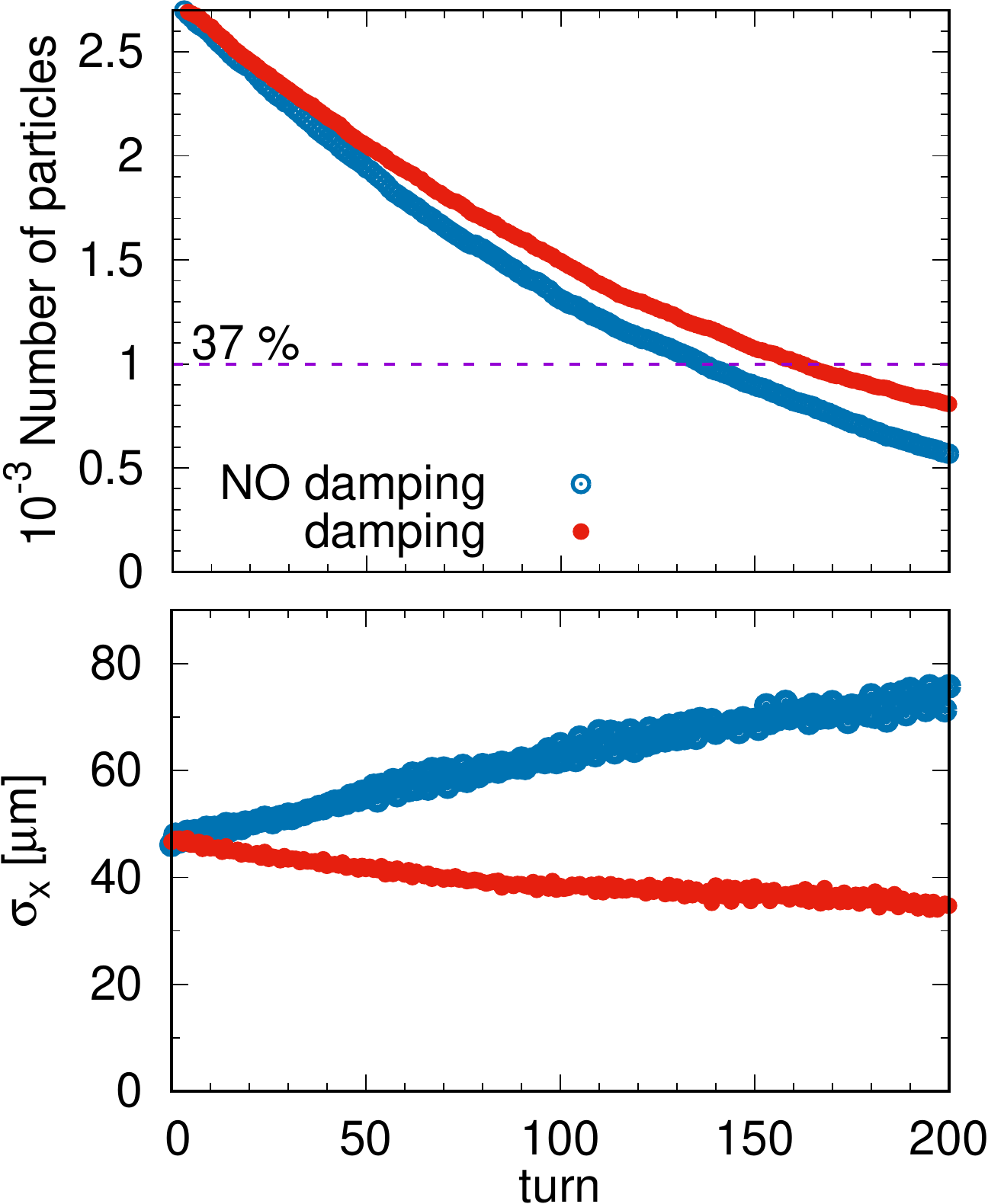}
   \caption{Beam lifetime (top) and horizontal beam size (bottom) with and without the damping effect in the tracking simulation including a 0.5~mm thick Be target.}
   \label{damping}
\end{figure}

Beam lifetime slightly increases due to radiation  damping. For a 3~mm Be target the effect is negligible because the ring longitudinal damping time is approximately twice the 40 turns we have shown, and the transverse damping time is almost four times larger. With a Be target of 500~$\mu$m the increase in beam lifetime amounts to 8\% of the 140 turns obtained from simulations with MAD-X, is  as shown in Figure~\ref{damping}.\par
\begin{figure}[bht]
   \centering
   \includegraphics*[height=8.cm,angle=-90]{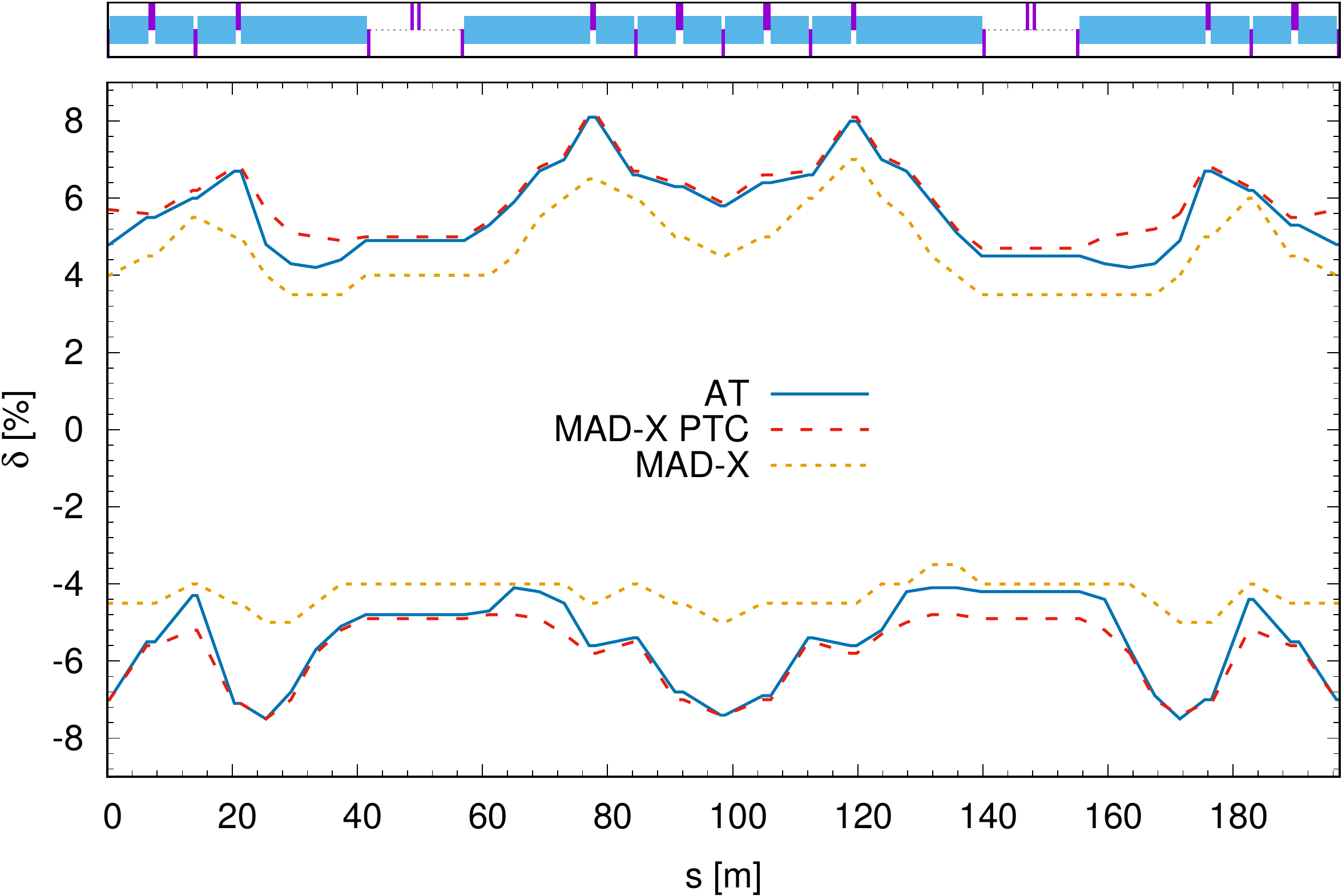}
   \caption{Momentum acceptance of the positron ring cell obtained from tracking with MAD-X (dotted yellow), MAD-X PTC (dotted red) and AT (light blue). 
   The best agreement is obtained between AT and MAD-X PTC.}
   \label{moma}
\end{figure}

Lifetime is limited by the energy loss, thus it is important to maximise the ring momentum acceptance. Figure~\ref{moma} shows that the cell without errors reaches 8\% of momentum acceptance, obtained from particle tracking along the ring with three different tracking codes: AT, MAD-X and MAD-X~PTC.  Good agreement is found between the codes.\par
\par
\begin{figure}[!htb]
   \centering
   \includegraphics*[width=8.5cm]{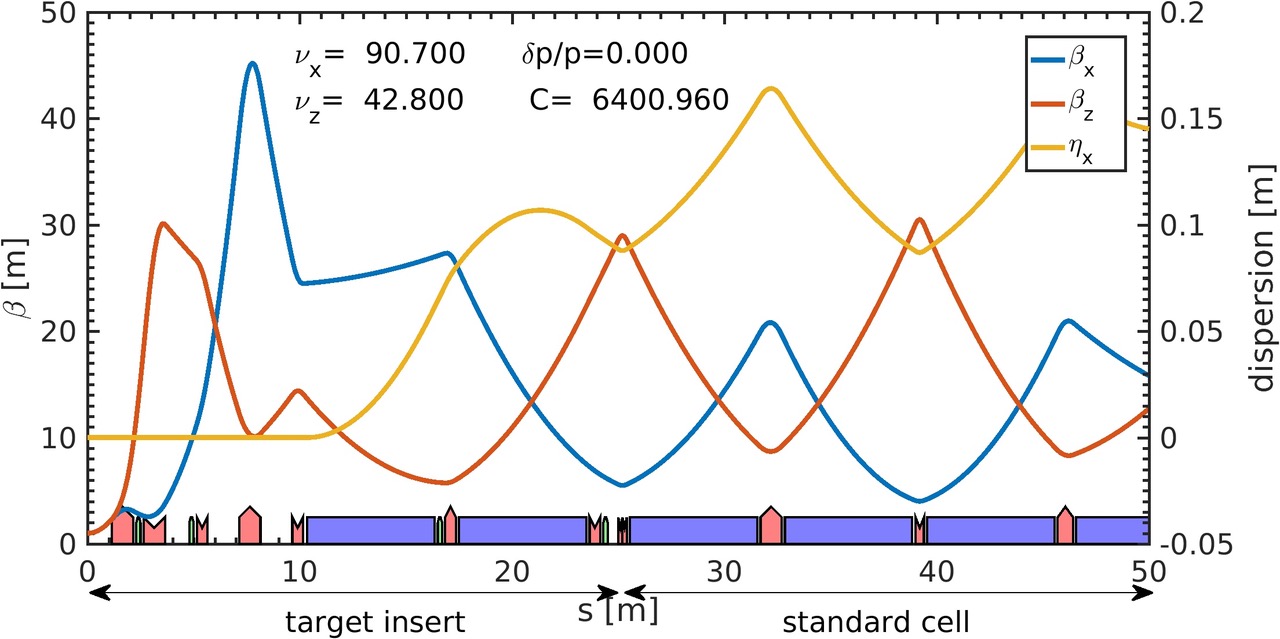}
   \caption{Target insertion region (target is at s=0~m) and 25~m of the cell. $\beta_{x}$ and $\beta_{y}$ functions are shown in blue and  red, respectively; horizontal dispersion function $\eta_{x}$ is plotted in yellow with  values  on the right y axis.}
   \label{IRcell}
\end{figure}

One of the ring cells has been modified to obtain a low-$\beta$  interaction region to place the target at its center similarly to the interaction point  of a collider.
Figure~\ref{IRcell} shows the optics and layout of the current target insertion region and 25~m of the cell. At the target location, $s=0$~m, the optical functions are $\beta_{x}=\beta_{y}=0.5$~m and $\eta_{x}=0$~mm.
\par
\begin{figure}[bht]
   \centering
   \includegraphics*[height=8.5cm,angle=-90]{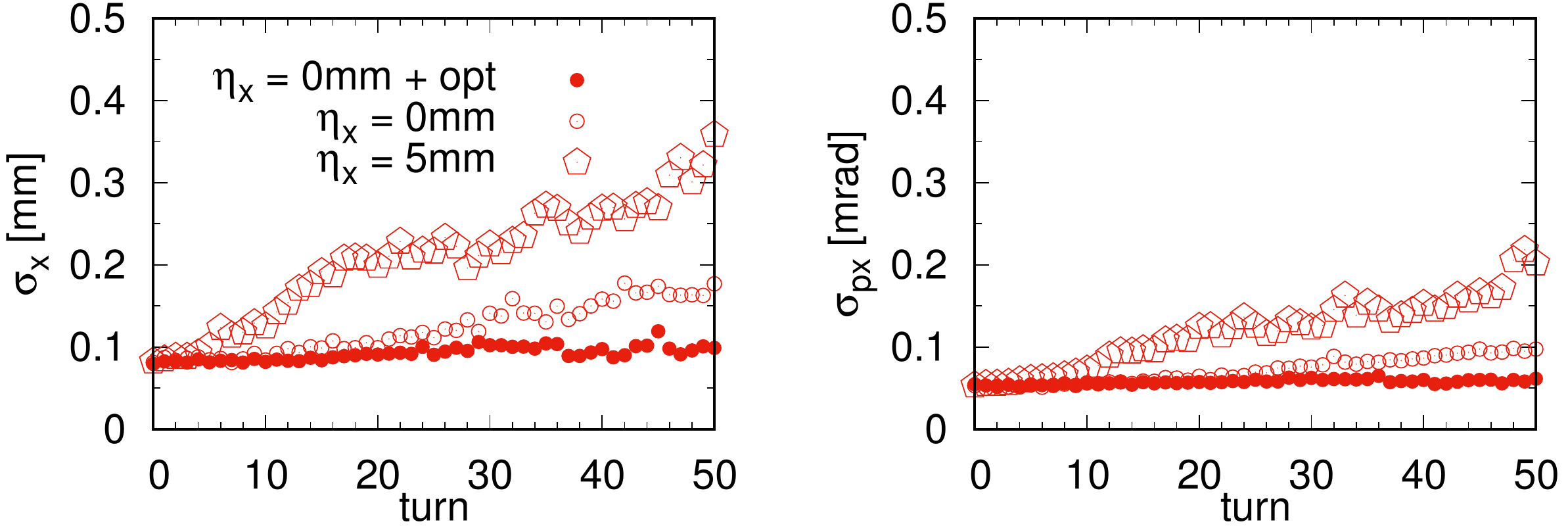}
   \caption{Effect of the cancellation of the horizontal dispersion at the target location and optimization of the lattice on the beam size growth and divergence from bremsstrahlung. Full red dot markers show the 
   best case with no degradation of $\sigma_{x}$ and $\sigma_{x'}$, where linear and nonlinear dispersion  terms at target are set to zero.}
   \label{radiative}
\end{figure}

In addition, linear and non-linear terms related to the momentum deviation, $\delta p$, should be minimized because they contribute to the emittance growth.
Figure~\ref{radiative} shows the positron beam size at the target increasing  as a function of the machine turns for different values of dispersion function in the target insertion  region, and  with $\beta$ functions unchanged.
When the dispersion at the target is cancelled, the beam size increase due to target interactions is damped.\par

\par
Figure~\ref{horbeam} shows the results of beam emittance degradation with the 
contributions from multiple scattering and bremsstrahlung separated. 
Values have been calculated using the 95~\% core  of the particle distribution.
The current design status amounts only to a two-fold horizontal emittance increase and a four-fold in vertical by the end of the lifetime.
The Bremsstrahlung energy loss in the target gives a longitudinal emittance growth. This increase of emittance and corresponding bunch lengthening can be reduced by decreasing the ring momentum compaction and increasing the RF voltage.  

 The longitudinal emittance growth is given completely by the energy loss per turn making the beam longer.\par

\begin{figure}[!htb]
   \centering
   \includegraphics*[width=250pt]{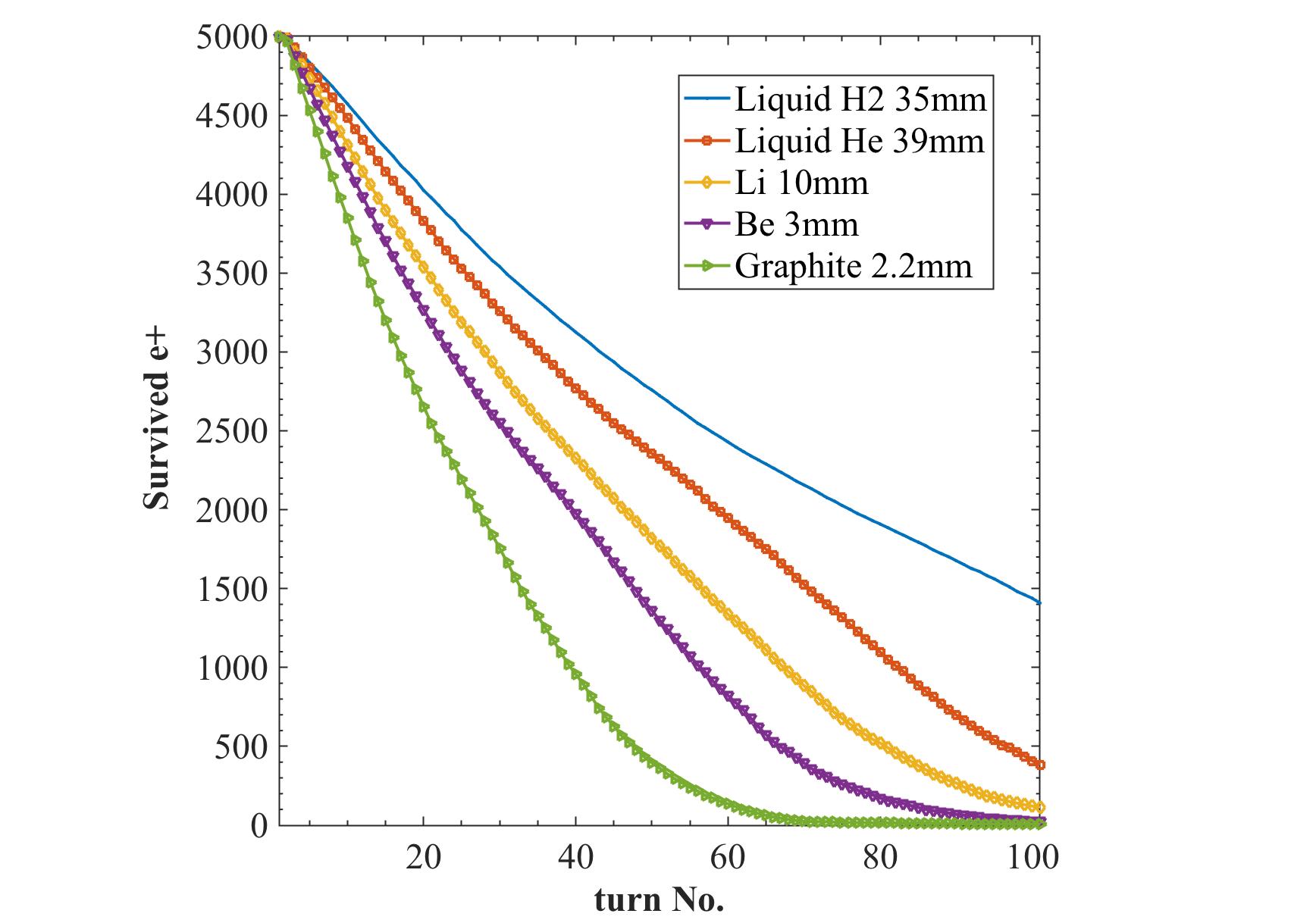}
   \caption{Number of e$^+$  vs turns number for different target materials. Target thickness has been chosen to obtain a constant muon yield. 3~mm Be, shown in this plot in dark violet, is the case studied in our beam dynamics simulations.}
   \label{lifetimemat}
\end{figure}

Thin  low Z targets ({\it i.e.} thickness of the order of 0.01~Xo)  based on Li, Be and C could be used for low emittance muon production. 
For equivalent electron density in the target, lighter materials will provide smaller beam perturbations at the cost of larger intrinsic muon beam emittances. 
Figure~\ref{lifetimemat} shows  the number of survived positrons as a function of machine turns
for different material targets. A 10~mm Lithium target might provide sizeably larger lifetime at the cost of a factor three increase in the intrinsic muon beam emittance.
To maximize the brillance of the muon beam the positron beam spot at the target has
 to be minimized and the positron beam intensity  maximized.

\section{Target consideration}
\label{target}
Both temperature rise and thermal shock are related to the beam size on target. 
For a given material the lower limit on the beam size is obtained when there 
is no pile-up of bunches on the same target position.
For this reason both the target and the positron beam have to be movable.
A fast beam bump  can be done after the extraction of one muon bunch from the muon accumulator ring every one muon laboratory lifetime, corresponding to  approximately 2500 positron bunches.
Fast moving targets can be obtained with rotating disks for solid targets or high velocity jets for liquids. 
A beam current of about 200~mA  will provide  about 100~kW of power that has to be removed to keep the target temperature under control.  
Be and C composites/structures are in use and under study for low Z target and collimators in accelerators for high energy physics also because of the stringent vacuum requirements in such complexes that are not easy to fulfil with Li targets.
Recently developed C based materials with excellent thermo-mechanical properties are under study for the LHC upgrade collimators~\cite{ipacT}. 
A 7.5 $\mu$s long beam pulse made of  288 bunches with 1.2$\times$10$^{11}$ protons per bunch, which is the full LHC injection batch extracted from SPS, has been used to test both C-based~\cite{ipacT} and Be-based~\cite{targetry} targets with maximum temperatures reaching 1000$^\circ$~C. 
Good results have been obtained with a beam spot of $0.3\times0.3$~mm$^{2}$.
Liquid Lithium, LLi, is under study for  divertors in tokamaks and in use for neutron production. In particular, LLi jets in the order of 1~cm thickness  are used as targets for  MeV proton beams to produce high flux neutron beams. Power deposited on target of  about 0.5~MW and of $>$ 1~MW have been obtained or planned with LLi jet velocities of 5~m/s or 30~m/s respectively~\cite{LLi1}.  In addition to  evaporation,  beam spot sizes smaller than 1 mm$^2$ would reach the bubbling regime for LLi. 
One possibility to overcome this problem is to produce a fast LLi  jet in a thin pipe. 

\section{Positron source requirements}
\label{posiso}
\begin{table*}[hb]
    \centering
  \caption{Positron sources parameters for future projects from ref.~\cite{Yakimenko:2012zzb}. \label{tab:Esources}. The values given for LEMMA  refer to the goal performances.}
  \vspace{0.05cm}
  \small
  \begin{tabular}{*{7}{c}}
    \hline\hline
    \noalign{\vspace{0.05cm}}
      & SLC & CLIC & ILC & LHeC & LHeC ERL& LEMMA goal\\
    \hline
    E [GeV] & 1.19 & 2.86 & 4 & 140 & 60 &45 \\
    $\gamma\epsilon_{x}$ [$\mu$m] & 30 & 0.66 & 10 & 100 & 50   &0.04   \\
    $\gamma\epsilon_{y}$ [$\mu$m] & 2  & 0.02 & 0.04 & 100 & 50 & 0.04\\
    $e^{+}$[$10^{14}$s$^{-1}$]    & 0.06 & 1.1 & 3.9 & 18 & 440 & 100\\
    \hline\hline
  \end{tabular}
  \end{table*}

The scheme described in the previous section relies on the possibility of enhancing the muon production by recirculating the positron beam, 
allowing multiple beam target interactions.
However, every time the primary beam interacts in the target, positrons will lose part of their energy, and a certain fraction of them will exceed the energy acceptance of the machine, being eventually lost. An high intensity  positron source is needed to replace these losses. 

 The present record positron production rate has been reached at the SLAC linac SLC. A summary
of the parameters of the positron sources for the future facilities
is reported in Table ~\ref{tab:Esources}. The ILC positron source has been designed to
provide $3.9\times 10^{14} e^+/$s. Two order of magnitudes more intense sources are foreseen for LHeC. LHeC is an electron machine 
although a positron option has been conceived.

The required intensity is strongly related to the beam lifetime that is determined by the ring momentum acceptance and the target material, see sections~\ref{optics} and \ref{target}.
The  beam lifetime with the present optics (energy acceptance about 6\%)  is in the range of 40-50 turns corresponding to about 2-3\% of positrons  lost per turn. 
By increasing the ring circumference we aim at increasing the energy acceptance.\\
With 50 turns lifetime, the scenario requires $3 \times 10^{16}$~45~GeV $e^+$/s . This corresponds to a beam power of 216~MW.
A lifetime of 100 turns can be obtained with an H pellet target and energy acceptance of about 10\%.

In the interaction of the primary positron beam with the Beryllium target, Bremsstrahlung photons are produced with a strong boost along the primary beam direction. It may then possible to exploit this photon flux for an embedded positron source. A thick high Z target placed downstream of the muon target
 can be used for electron positron  pair production.    
 
Experimental tests of such a scheme based on the adiabatic matching device collection scheme have been performed at KEK~\cite{Chaikovska:2017amy}.

However we do not yet have a system that is able to transform the temporal structure of the produced positrons to one that is compatible
with the requirement of a standard positron injection chain.
To evaluate the performance of such a scheme, a full Monte Carlo simulation has been performed, using GEANT4 for the part related to the positron production.
 
\begin{figure*}[!hb] 
 \begin{center} 
\includegraphics [width=10cm,angle=90] {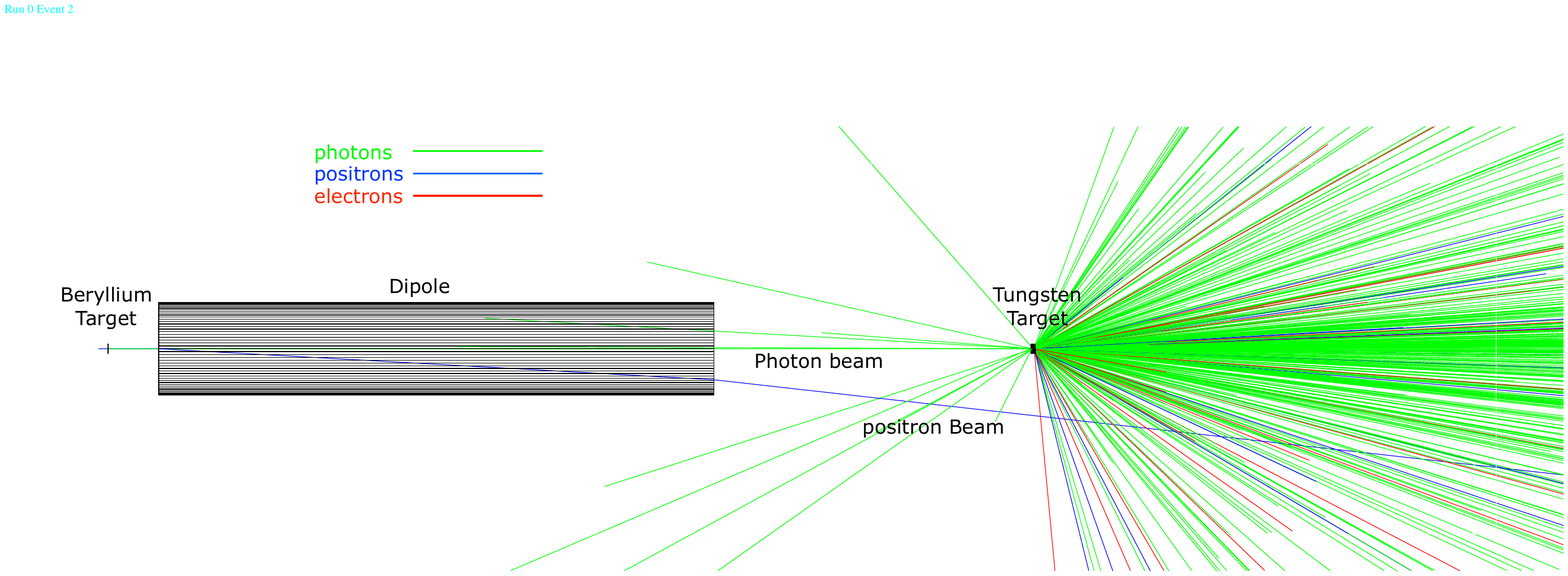} 
\caption{\small{Event display of GEANT4 for positron production.}} 
\label{fig:scheme} 
\end{center} 
\end{figure*}

A simulation of the target system has been performed with GEANT4. 
The geometry set-up is shown in Figure~\ref{fig:scheme}. A 45~GeV monochromatic positron beam is sent towards a 3~mm Beryllium target. After the target, a dipole magnet of the positron ring bends away charged particles, while the produced photons proceed straight, interacting with the Tungsten target, where $\rm e^+ e^-$ pairs are created.

It has been found that  in the interaction of $100$ positrons within the Beryllium target, about $11$ photons and $5$ electrons are created on average.
 As expected, photons are produced with a Bremsstrahlung energy spectrum  and strongly boosted along the beam axis.
\begin{figure}[!ht] 
\begin{center} 
  \includegraphics [height=5.cm] {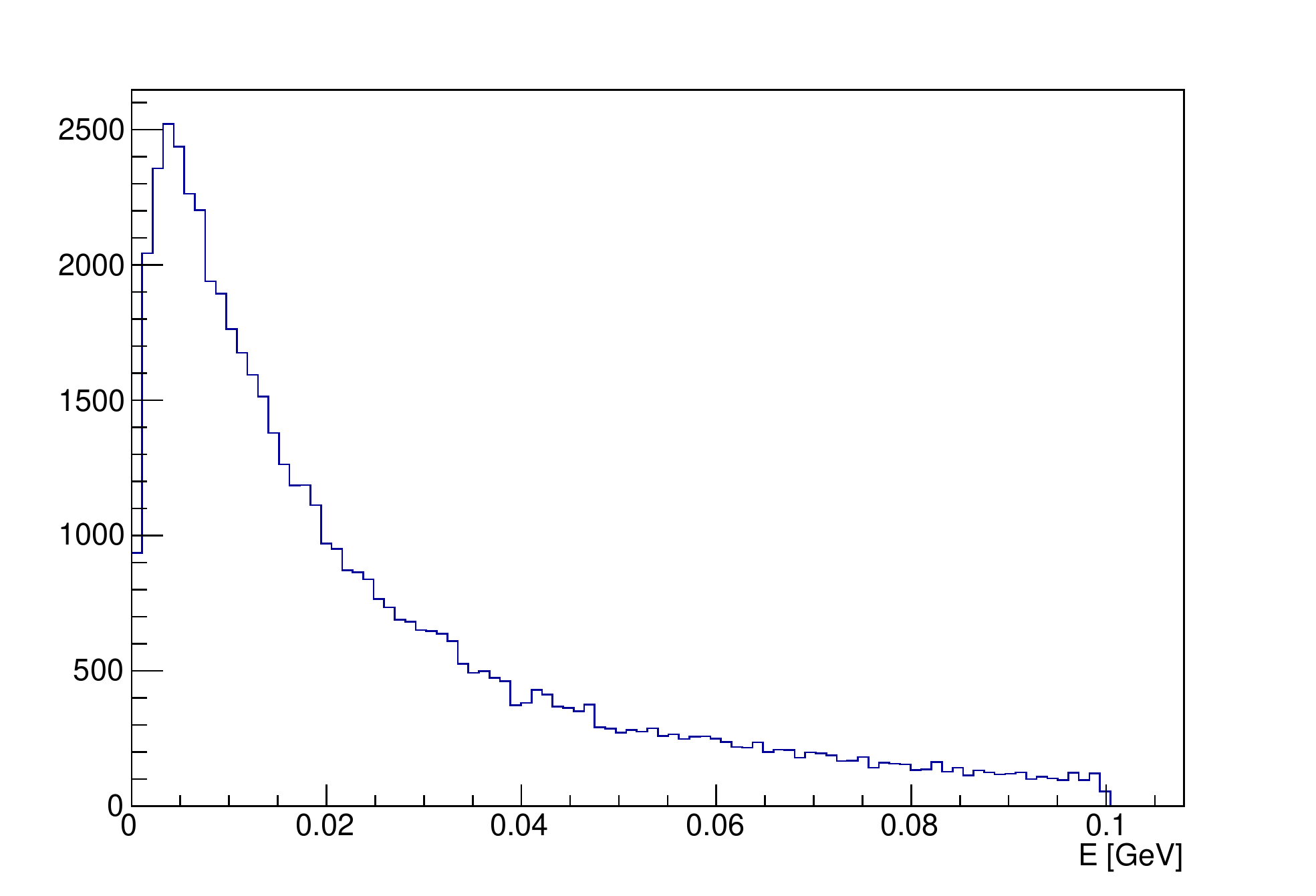} 
  \includegraphics [height=5.cm] {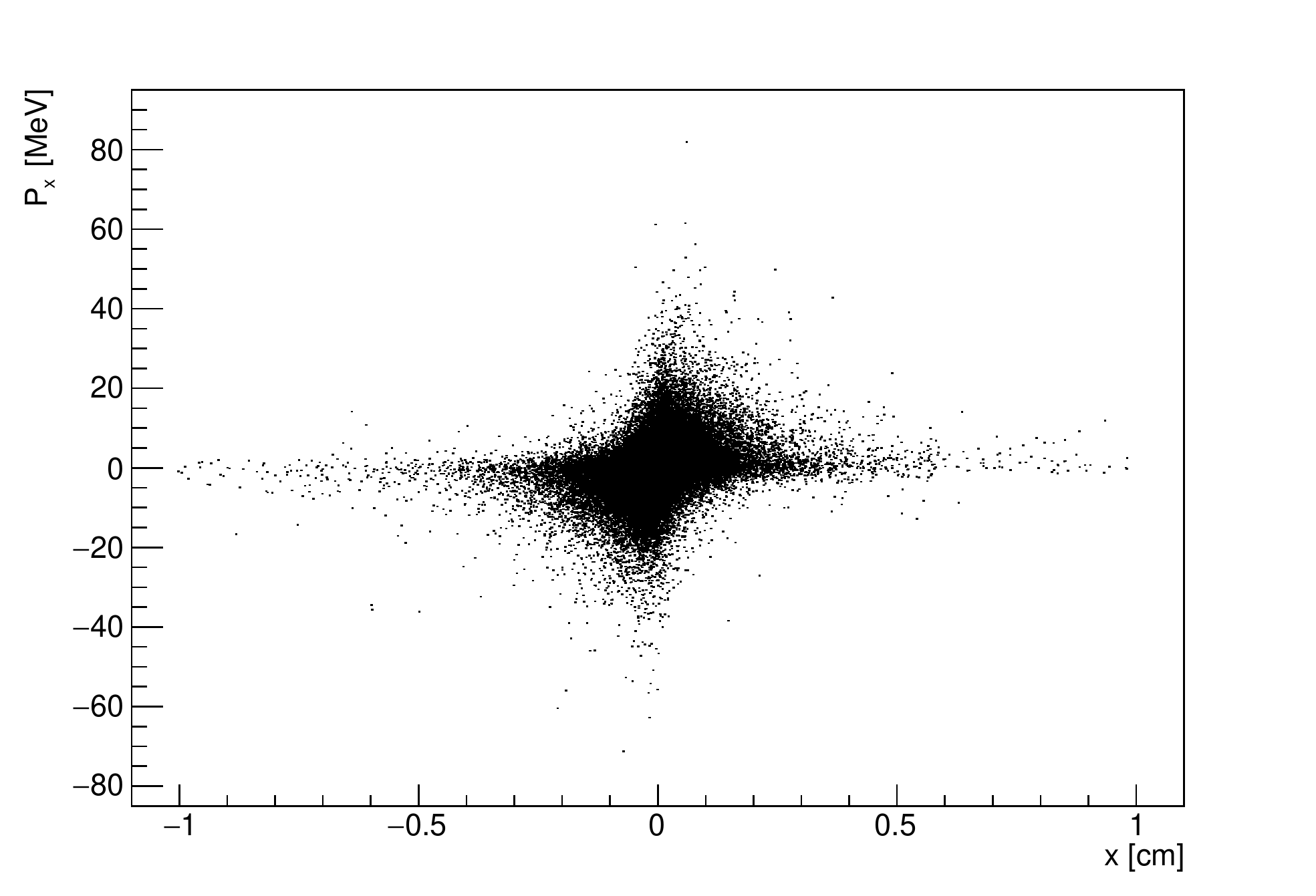} 
\caption{\small{Positrons energy spectra and  transverse phase space distribution  after the Tungsten target (GEANT4)  from hitting photons produced at the muon target.}} 
\label{fig:produced} 
\end{center} 
\end{figure}
Only  photons with high boost  will reach the thick Tungsten target, with an average overall  yield of about 10.5 photons per 100 primary positrons, with a very small ($<0.1$ per mill) contamination of residual $\rm e^+/e^-$.
In the interaction of these 10.5 photons in a $\rm 5~X_0$ Tungsten target, about $65$ positrons are produced, together with 77 electrons and 1200 photons. Figure~\ref{fig:produced} shows the positron energy spectra and the positron transverse phase space distribution.

In conclusion, the suggested scheme  produces about 60 secondary positrons from the interaction of 100 primary positrons, of which  less than $\rm \sim 3$ are lost 
due to the interaction in the beryllium target. It would thus be enough to have a collection efficiency for these secondary positrons of about $5 \%$ in order to be able to recover the loss in the primary beam.
A full simulation in GEANT4 and ASTRA~\cite{astra} will be performed in future study, following the adiabatic matching scheme proposed in~\cite{Chehab:2000xc}.

\section{Conclusion and Perspectives}
We have presented a novel scheme for the production of muons starting
from a positron beam on target, discussing some of the critical aspects and
key parameters of this idea and giving a consistent set of possible
parameters to show its feasibility. This scheme has several
advantages, the most important one is that it can provide low emittance muons without adding cooling.

This innovative scheme has many key topics to be investigated: a low emittance 45 GeV positron ring,
O(100~kW) class target,  high rate positron source, and a 22 GeV muon accumulator ring.

We presented the preliminary study of a 45 GeV  positron ring with a thin Beryllium target insertion.
The ring has an high momentum acceptance allowing a lifetime of about 40 turns for a 3 mm Be target.
 Beam emittance growth due to the interaction with target has been observed. A dedicated cell has been designed to show that 
the emittance growth can be contained with proper optics parameters at the target location.

We have shown that this effect can be reduced  by lowering the value of the $\beta$-function at the target location in addition to the minimization of the linear and high order dispersion terms.
However stringent constraints on the beam size at the target are imposed by thermo-mechanical stresses, posing a lower limit on the emittance that can be obtained.
Although an increase in emittance with respect to that shown in Figure~\ref{f2} seems unavoidable, it might be compensated with an increase of the positron beam energy
up to 50~GeV where a factor of two in the muon production rate can be obtained, at the cost of more than a factor four larger emittance with almost no loss in muon collider luminosity performance.
Nevertheless, one has to cope with a much larger muon energy spread, up to 18\%.

Further improvements in nonlinear effects corrections to increase the energy acceptance of the positron ring can be studied.
We need also to reduce the momentum compaction to be as close to zero as possible in order to decrease the 
positron -and thus also the muon- bunch length.  Beam instabilities driven by this small momentum compaction need to be studied, as well.\\
The possibility of increasing the ring circumference has to be fully explored since it would allow the reduction of the synchrotron radiation power loss. 
Keeping the same number of positrons per bunch and the same bunch distance one would get the same muon production rate for the same positron source requirements.
 In addition, by increasing the ring circumference, the ring parameters could be improved, reducing transverse emittance and momentum compaction and increasing dynamical aperture and momentum acceptance.
 Progress need to include all other topics like target material, muon accumulation issues, positron source and injection.

\section{Acknowledgments}
The authors thank L.~Keller for useful discussions and suggestions. The authors also thank I.~Chaikovska and R.~Chehab for discussions on the positron source and AMD scheme, 
S.~Gilardoni and M.~Calviani for suggestions about the target issues and H.~Burkhardt for discussions about GEANT4.

\bibliographystyle{unsrt}
%
%

\end{document}